\documentstyle{article}
  \def\da{\dagger}
  \def\disp{\displaystyle}

  \def\hA{\hat{A}}
  \def\hH{\hat{H}}
  \def\bq{{\mbox {\boldmath $q$}}}
  \def\bQ{{\mbox {\boldmath $Q$}}}

\title{\large{\bf{Unified Description of $\bq$-deformed Harmonic Oscillators}}}

\author{Ikuo S. {\sc Sogami}\footnote{E-mail: sogami@cc.kyoto-su.ac.jp} and Kouzou {\sc Koizumi}\footnote{E-mail: kkoizumi@cc.kyoto-su.ac.jp}}

\begin{document}

\maketitle
{\it Department of Physics, Kyoto Sangyo University,
Kyoto 603-8555, Japan}

\begin{abstract}
A wide class of $q$-deformed harmonic oscillators including those of 
Macfarlane type and of Dubna type is shown to be describable in a unified
way. The Hamiltonian of the oscillator is assumed to be given by
a $q$-deformed anti-commutator of the $q$-deformed ladder operators.
By solving $q$-difference equations, explicit coordinate representations
of ladder operators and wave functions are derived, and unified parametric
representations are found for $q$-Hermite functions and related formulas
for the oscillators of Macfarlane and Dubna types. In addition to the
well-known solutions with globally periodic structure, there exist
an infinite number of solutions with globally aperiodic structure.
\end{abstract}

\section{Introduction}
A prototype of $q$-deformed harmonic oscillator is found in the model which
was investigated by Macfarlane in a \lq\lq parametric coordinate\rq\rq\,
representation\cite{1}. In his original theory, the Hilbert space is
composed of wave functions related to the Rogers-Szeg\"{o} polynomials on
the unit circle\cite{2,3,4}. The $q$-deformed oscillators with a realistic
coordinate on the infinite interval were explored in succeeding
models\cite{5,6}. Independently of this lineage, Dubna group found
a different kind of $q$-deformed harmonic oscillators in the course of
their study of one dimensional relativistic system\cite{7}-\cite{13}.
In their approach also,
 a rather \lq\lq abstract coordinate\rq\rq\,
was used for the oscillator.

In this article, we investigate both types of $q$-deformed harmonic oscillators,
those of Macfarlane type and of Dubna type, in a common realistic coordinate
representation and find a wide class of new oscillators as their descendants.
As a basic postulate, ladder operators of the oscillator are assumed to be
composed of products of such part-functions that include separately the
coordinate operator and the momentum operator of the oscillator\cite{7,10,11}.
In particular, a kind of $q$-deformed derivative consisting of a difference of
exponetial functions is chosen for the part-function of the momentum.
Other part-functions are obtained by solving $q$-difference equations deduced
from the condition that the ladder operators should satisfy a $q$-deformed
commutation relation.

One of the part-functions of coordinate is obtained as an infinite sum of
Gaussian functions with arbitrary coefficients. It is the freedom in
the arbitrariness of coefficients that enables us to determine global
structure of the $q$-deformed oscillator system. Connection of the ladder
operators and wave functions by adjusting properly the coefficients
results in the periodic global structure, which was recognized first
in the oscillators of Macfarlane type\cite{5,6}. In addition to such
globally periodic solution, there exist new infinite solutions with
aperiodic global structure.

The Hamiltonian which is assumed to be given by a $q$-deformed
anticommutator of the ladder operators has a common eigenvalue spectrum,
expressed by a definite function of the deformation parameter, for
all $q$-deformed oscillators studied in this article. Derivation of
its eigenfunctions results in different $q$-deformed Hermite
functions\cite{5,10,14}, respectively, for the oscillators of
Macfarlane and Dubna types.

In this constructive approach, it turns out to be possible to clarify
similarity and difference of the $q$-deformed oscillators of Macfarlane and
Dubna types in a unified way and to disclose explicitly the existance of new
infinite families of oscillator systems with different global structure.

\section{Algebra of $\bq$-deformed harmonic oscillator}
Let us investigate a system of $q$-deformed harmonic oscillator which has
a deformation parameter $q$ and the Hamiltonian $\hH_q$
\begin{eqnarray}
     \hH_{q}=\frac{1}{2}\{\hA, \hA^{\dag}\}_{q}
     =\frac{1}{2}(q\hA\hA^{\dag}+q^{-1}\hA^{\dag}\hA),
     \label{H_q}
\end{eqnarray}
where $\hA$ and $\hA^\dag$ are, respectively, raising and lowering operators
satisfying the $q$-deformed commutation relation ($q$-mutator):
\begin{eqnarray}
   [\hA, \hA^{\dag}]_{q}=q\hA\hA^{\dag}-q^{-1}\hA^{\dag}\hA = 1.
   \label{q-mutator}
\end{eqnarray}

As in the case of  the ordinary (non-deformed) harmonic oscillator system,
the ground state for the Hamiltonian $\hH_q$ is defined by the conditions
\begin{eqnarray}
     \hA|0\rangle=0,\quad \langle 0|0 \rangle = 1
\end{eqnarray}
and excited states are generated from it by applying the raising operator as
\begin{eqnarray}
     |n\rangle=N_n(\hA^{\dag})^n|0\rangle.
     \label{nstate}
\end{eqnarray}
To confirm that $|n\rangle$ are eigenstates of the Hamiltonian and to obtain
the eigenvalue, it is sufficient to use the following $q$-deformed commutation
relation
\begin{equation}
     [\hA, \hH_q]_q = \frac{1}{2}(q+q^{-1})\hA,
\end{equation}
which leads readily to the recursion formula
\begin{equation}
    q\left(E_n - \frac{1}{2}\frac{q+q^{-1}}{q-q^{-1}} \right)
    = q^{-1}\left(E_{n-1} - \frac{1}{2}\frac{q+q^{-1}}{q-q^{-1}}\right).
\end{equation}
Using this relation and noting $E_0=\frac{1}{2}$, we find the general
formula for the energy eigenvalue spectrum of the $q$-deformed harmonic
oscillator as follows:
\begin{eqnarray}
    E_n(q) = \frac{\ \ 1-q^{-2n}}{q^2-1} + \frac{1}{2}.
    \label{eigenvaluespec}
\end{eqnarray}
The equal spacing law of eigenvalues of the ordinary harmonic oscillator
is deformed into the geometrical progression
\begin{equation}
    E_{n+1} - E_n = q^{-2}\left( E_n - E_{n-1} \right),
    \qquad n = 1, 2, \cdots
\end{equation}
with constant ratio $q^{-2}$ for the difference of adjacent eigenvalues.
Note that this spectrum is derived by algebraic procedure only. Therefore,
its form is common to all $q$-deformed oscillators specified in later sections
and is independent of details of their representations. Normalization of
the eigenvectors results in the recursion formula
\begin{equation}
    \left(\frac{N_n}{N_{n+1}}\right)^2 = \langle n|\hA\hA^\dag|n\rangle
    = \frac{1}{2q}\left(2E_n+1\right),
\end{equation}
from which the normalization constant $N_n$ in Eq.~(\ref{nstate}) is
derived as
\begin{equation}
    N_n=\prod_{m=1}^{n}\left(\frac{q-q^{-1}}{1-q^{-2m}}\right)^{\frac{1}{2}}.
\end{equation}

\section{$\bQ$-deformed ladder operators in $x$-representation}
Following the basic postulate made in the Introduction, the part-function
of the ladder operators which includes the momentum operator $\hat{p}$ is
assumed to take the form
\begin{equation}
  D(\hat{p})=\frac{i}{(s-t)}\left[\exp(s\hat{p})-\exp(t\hat{p})\right]
\end{equation}
where $s$ and $t$ are real parameters. Note that this part-function is 
a kind of $q$-deformed derivative which has the limit
\begin{equation}
    \lim_{s\rightarrow 0}\lim_{t\rightarrow 0}D(\hat{p}) =
    \lim_{t\rightarrow 0}\lim_{s\rightarrow 0}D(\hat{p}) =
    i\hat{p}.
\end{equation}
Other part-functions depending on the coordinate operator $\hat{x}$ are
introduced below at the stage of taking explicit coordinate representation
and determined by solving $q$-difference equations. It turns out to be relevant
to define the deformation parameter $q$ by
\begin{equation}
    q=\exp\left( s^2 + t^2 + 3st \right)
\end{equation}
in terms of the real parameters $s$ and $t$.

For an arbitrary state vector $|\psi\rangle$, the ladder operators $\hA$ and
$\hA^\da$ consisting of separable part-functions have $x$-representations
defined by
\begin{equation}
   \langle x|\hA|\psi\rangle = A(x)\psi(x),\quad
   \langle x|\hA^\da |\psi\rangle = A^\da(x)\psi(x)
\end{equation}
where $\psi(x)=\langle x|\psi\rangle$. The raising and lowering operators
in the $x$-representation, $A(x)$ and $A(x)^\da$, are postulated to have
the following separable forms
\begin{eqnarray}
   A(x) = \frac{f(x)}{g(x)}\exp[-i h(x)]
          D\left(\frac{1}{i}\frac{d}{d x}\right)
          \frac{1}{f(x)g(x)}
   \label{A}
\end{eqnarray}
and
\begin{eqnarray}
    A^{\dag}(x) = -\frac{1}{f(x)g(x)}
                  D\left(\frac{1}{i}\frac{d}{d x}\right)
                  \frac{f(x)}{g(x)}\exp[i h(x)]
   \label{Adag}
\end{eqnarray}
where $f(x)$, $g(x)^2$ and $h(x)$ are functions which take real values for
$x \in {\bf R}$. It is essential to assume that the functions $f(x)$, $g(x)$
and $h(x)$ are continued analytically into the complex $x$ plane. These
functions are determined from the conditions that the ladder operators satisfy
the $q$-mutator in Eq.~(\ref{q-mutator}) and reduce to those of the
ordinary (non-deformed) hamonic oscillator at the limit $q\rightarrow 1$.
By definition, the function $f(x)$ has intrinsic uncertainty of
an arbitrary multiplicative constant and the sign of the function $g(x)$
also is indeterminate. As will be clarified in sections 5 and 6 and in
the Appendix, the function $g(x)$ turns out to have further intrinsic
freedoms of uncertainty in the $q$-deformed oscillator systems of Dubna type.

As a necessary condition that the basic $q$-mutator in Eq.~(\ref{q-mutator})
includes a constant term, the parameters $s$ and $t$ must satisfy
one of the following conditions
\begin{equation}
      s = 0\ \ {\rm or}\ \ t = 0
   \label{t=0}
\end{equation}
and
\begin{equation}
      s + t = 0.
   \label{t=-s}
\end{equation}
Without loss of generality, we select out here the parameter $s$ as a basic
one. Namely, the parameter $t$ is elliminated by choosing the condition
$t = 0$ in Eq.~(\ref{t=0}) and by setting $t=-s$ in Eq.~(\ref{t=-s}).
As shown in subsequent sectoins, the systems being subject to the former and
latter conditions are identified generically with the $q$-oscillators of
Macfarlane type and of Dubna type, respectively.

The $x$-representation of the Hamiltonian $\hat{H}_q$, $H_q$, is
constructed as follows:
\begin{equation}
   \langle x|\hat{H}_q|\psi\rangle = H_q\psi(x) =
   \frac{1}{2}\left[qA(x)A^\da(x)+q^{-1}A^\da(x)A(x)\right]\psi(x).
\end{equation}
The ground-state eigenfunction $\psi_0(x)$ of the Hamiltonian
$H_q$ which is annihilated by the lowering operator as
$A(x)\psi_0(x)=\langle x|\hat{A}|0\rangle=0$ is determined to be
\begin{eqnarray}
     \psi_0(x)=\langle x|0\rangle=\disp K_0(I) f(x)g(x)
     \label{groundstate}
\end{eqnarray}
where the normalizatoin constant $K_0(I)$ is given by
\begin{eqnarray}
    K_0(I) = \disp{\left[\int_I f(x)^2 g(x)^2 dx\right]^{-\frac{1}{2}}}.
\end{eqnarray}
For wave functions $\psi(x)$ and $\phi(x)$, the inner product is defined by
\begin{equation}
    \langle\psi|\phi\rangle
    =\disp{\int_I dx\,\langle\psi|x\rangle\langle x|\phi\rangle}
    =\disp{\int_I dx\,\psi^{\ast}(x)\phi(x)}
\end{equation}
where $I$ is the domain for integration which depends on global structure of
the $q$-deformed oscillators specified in section 6. It is naturally required
to prove later that the operators $A(x)$ and $A^\dag(x)$ are mutually adjoint
on the Hilbert space generated by eigenstates of the Hamiltonian $H_q$.

In the following analysis, it is convenient to introduce the function
\begin{eqnarray}
    F(x)=\disp\left[\frac{f(x+i s)}{f(x)}\right]^2.
    \label{F(x)}
\end{eqnarray}
Note that, owing to the analyticity of the function $f(x)$, the function
$F(x)$ must satisfy the condition $\lim_{s \rightarrow 0}F(x)=1$.

\section{$\bQ$-deformed harmonic oscillators of Macfarlane type 
($q=e^{s^2}$)}
Let us investigate first the $q$-deformed harmonic oscillators of
Macfarlane type. In this case ($t=0$) where the deformation
parameter is given by $q = e^{s^2} \geq 1$, there exist the upper
as well as lower bounds in the energy spectrum for every definite
$q$ except for $q\neq 1$. Namely
\begin{equation}
    \frac{1}{2} \leq  E_n < \frac{1}{2} + \frac{1}{q^2-1}.
\end{equation}
At the limit $s \rightarrow \infty$, all the eigenvalues accumulate onto
$\frac{1}{2}$. For the right-hand-side of the basic $q$-mutator to be
the constant 1, the functions $f(x)$, $g(x)$ and $h(x)$ must satisfy
the following three relations as
\begin{equation}
   \disp\frac{q-q^{-1}}{s^2 g(x)^4}=1,
   \label{macfarlanecond1}
\end{equation}         
\begin{eqnarray}
   &&\hspace{-1cm}\left[\frac{f(x-i s)}{f(x)g(x)^2}+\frac{f(x)}{f(x-i s)g(x-i s)^2}
   \right]
   \nonumber\\
   \noalign{\vskip 0.2cm}
   && = q^{-2}
   \left[\frac{f(x)}{f(x-i s)g(x)^2}+\frac{f(x-i s)}{f(x)g(x-i s)^2}
   \right]\exp\{i[h(x)-h(x-i s)]\}
   \label{macfarlanecond2}
\end{eqnarray}
and 
\begin{eqnarray}
   \disp f(x)^2f(x-2i s)^2=q^{-2}f(x-i s)^4\exp\{i[h(x)-h(x-2i s)]\}.
   \label{macfarlanecond3}
\end{eqnarray}
Taking the sign ambiguity of the function $g(x)$ into consideration, we
choose the solution of the equation (\ref{macfarlanecond1}) as
\begin{equation}
   g(x)=\disp\left(\frac{q-q^{-1}}{s^2}\right)^{\frac{1}{4}}.
   \label{gMac}
\end{equation}
The fact that $g(x)$ is a constant simplifies the equation
(\ref{macfarlanecond2}) into the form
\begin{equation}
   \exp\{i[h(x)-h(x-i s)]\} = q^{2}
   \label{phase-eq}
\end{equation}
which results in
\begin{equation}
      h(x)-h(x-i s)=-2i s^2 +2\pi l
\end{equation}
with an arbitrary integer $l$. This difference equation has a general
solution
\begin{equation}
 h(x) = -2sx - i\frac{2\pi l}{s}x 
  +\sum^{\infty}_{n=-\infty}a_n\exp{\left(\frac{2\pi n}{s}x\right)}
\end{equation}
where coefficients $a_n$ are arbitrary. Notice that, as a due consequence
of the difference equation, a periodic function of $x$ with the period
$i s$ appears here as a power series of scaled variable $x/s$ in the
solution $h(x)$. Owing to the condition that the function $h(x)$ must be
real for $x\in {\bf R}$, the integer $l$ must be specified to be 0.
Further, for the function $h(x)$ to be definite for $x\in {\bf R}$
at the limit $s\rightarrow 0$, $a_n = 0$ for $n \not= 0$. As a result,
we obtain
\begin{equation}
     h(x)= -2sx + a_0.
     \label{h(x)Mac}
\end{equation}

In terms of the function $F(x)$ in Eq.~(\ref{F(x)}), the relation
(\ref{macfarlanecond3}) is expressed by
\begin{equation}
   \disp F(x+i s)=q^2 F(x)
   \label{gma-def}
\end{equation}
which has a general solution
\begin{equation}
   F(x) = \left[\sum^{\infty}_{n=-\infty}b_{n}\exp\left(\frac{2n\pi}{s}x\right)
          \right]\exp(-2i sx).
   \label{F(x)sol}
\end{equation}
Here the periodic function with period $i s$ appears as a multiplicative
uncertainty. Owing to the necessary condition $\lim_{s\rightarrow 0}F(x)=1$,
the coefficients in the periodic function are severely restricted to be
$b_0 \neq 0$ and $b_{n}=0\ (n\not=0)$. Consequently, we obtain
\begin{eqnarray}
    F(x) \equiv \disp\left[\frac{f(x+i s)}{f(x)}\right]^2
         = q\exp(-2i sx) = \exp(s^2 -2i sx).
    \label{F(x)solution}
\end{eqnarray}
The coefficient $b_0=q$ is specified from the definite form of
the function $f(x)$ derived below. Then, solving the equation (\ref{F(x)sol})
with respect to the function $f(x)$, we are able to determine the function
$f(x)$ which takes real values for $x\in {\bf R}$ in the following form
\begin{eqnarray}
     f(x) &\!\!\!=&\!\!\! \sum^{\infty}_{m=-\infty}c_m
     \exp\left(-\frac{s^2}{8}-\frac{2m^2\pi^2}{s^2}\right)
     \exp\left(\frac{2m\pi}{s}x\right)
     \exp\left[-\frac{1}{2}\left(x-i\frac{1}{2}s\right)^2-i\frac{1}{2}sx\right]
     \nonumber\\
     \noalign{\vskip 0.2cm} 
     &\!\!\!=&\!\!\! \sum^{\infty}_{m=-\infty}c_m
     \exp\left[-\frac{1}{2}\left(x-\frac{2m\pi}{s}\right)^2\right]
 \label{f(x)solution}
\end{eqnarray}
where coefficients $c_m$ must be chosen so as to make the function $f(x)$
to be square integrable, {i.e.}, 
\begin{equation}
    \int_{-\infty}^{\infty}f(x)^2 dx 
    = \sqrt{\pi}\sum_{n,m}c_n c_m
      \exp\left[-\frac{(m-n)^2\pi^2}{s^2}\right] < \infty.
\end{equation}
In the derivation of the part-function in Eq.~(\ref{f(x)solution}), it is
implicitly assumed that the central Gaussian component with $m=0$ in the sum
which survives at the limit $s \rightarrow 0$ is symmetric with respect to the
origin $x=0$.

In this way, all part-functions of the ladder operators $A(x)$ and $A^\dag(x)$
in Eqs.~(\ref{A}) and (\ref{Adag}) have been obtained. In the case of
$h(x)=-2sx$ where $a_0=0$ in Eq.~(\ref{h(x)Mac}), we find the solution
of Macfarlane type which was investigated by Shabanov\cite{5}.

In order to derive the eigenfunction of the $q$-Hamiltonian in equation
(\ref{H_q}), we utilize the relation among the ladder operators,
\begin{eqnarray}
  A^{\dag 2} = i\left(q-q^{-1}\right)^{-\frac{1}{2}}\exp[i h(x)]
           \left\{q^{-1}\exp(2i sx)-1\right\}A^{\dag}
   \nonumber\\
   \noalign{\vskip 0.2cm}
   -q^{-1}\exp\left\{2i[h(x)+sx]\right\}\left(q^{-2}A^{\dag}A + q^{-1}\right)
   \label{Ada2}
\end{eqnarray}
which leads readily to the recursion formula
\begin{eqnarray}
	\disp\psi_{n+1}(x)=\!\!\! &-i\left[1-q^{-2(n+1)}\right]^{-\frac{1}{2}}
     \exp[i h(x)]\left[q^{-1}\exp(2isx)-1\right]\psi_n(x)
    \nonumber\\
    \noalign{\vskip 0.2cm} 
  &-\exp\{2i[h(x)+sx]\}q^{-1}
      \left[\frac{1-q^{-2n}}{1-q^{-2(n+1)}}\right]^{\frac{1}{2}}\psi_{n-1}(x).
      \label{psirecursionMac}
\end{eqnarray}
To extract the $q$-deformed Hermite functions from the eigenfunctions,
which reduce properly to the Hermite polynomial at the limit $s \rightarrow 0$,
we set
\begin{eqnarray}
   \psi_n(x) = K_0\,f(x)g(x)&&\!\!\!\!\!\!\!\!\!\!\,s^n
           \exp\left\{i n[h(x)-sx]\right\}\nonumber\\
	\noalign{\vskip 0.2cm}
           &&\prod^{n-1}_{m=0}\left[q(1-q^{-2(m+1)})\right]^{-\frac{1}{2}}H_n(x;q^{-1}).
 \label{MacHermitedef}
\end{eqnarray}
provided that $\prod_{m=0}^{-1}1/\sqrt{q(1-q^{-2(m+1)})}=1$.
Then the equation (\ref{psirecursionMac}) gives rise to 
the recursion formula of the $q$-Hermite function $H_n(x;q)$ as
\begin{eqnarray}
  \disp H_{n+1}(x;q^{-1})
  = \frac{i}{s}
    \left[q^{-\frac{1}{2}}\exp(i sx)\right.&&\!\!\!\!\!\!\!\!\!\!\!\!\! \left.
    - q^{\frac{1}{2}}\exp(-i sx)\right] H_n(x;q^{-1})\nonumber \\
 \noalign{\vskip 0.2cm}
     &&-\disp\frac{1}{s^2}(1-q^{-2n})H_{n-1}(x;q^{-1}).
\end{eqnarray}
It is straightforward to prove that the relation
$N_n\,\psi_{n+1} = N_{n+1}A^\da\psi_n$ is equivalent to the second recursion
 formula of the $q$-Hermite function as follows:
\begin{eqnarray}
   && \hspace{-1.2cm}
      i s\left[q^{-\frac{1}{2}}\exp(i sx)+q^{\frac{1}{2}}\exp(-i sx)\right]
      H_{n+1}(x ; q^{-1})
      \nonumber \\
      \noalign{\vskip 0.2cm}
   && \hspace{-1cm}
      \disp{= q^{-n}\left[q^{-1}\exp(2i sx) H_n(x-i sx ; q^{-1})
      -q\exp(-2i sx)H_n(x+i sx ; q^{-1})\right]}.
    \label{Macsecondrecursion}
\end{eqnarray}

With these formulas and the conditions $H_0(x;q^{-1})=1$ and $H_{-1}(x;q^{-1})=0$,
the $q$-Hermite function is proved to have the power series representation
\begin{eqnarray}
  \hspace*{-0.5cm}
  H_n(x;q^{-1}) = \left(\frac{i}{s}\right)^{n}\sum^{n}_{m=0}
  (-1)^{m}q^{-\frac{(2m-n)}{2}}
   \left[\begin{array}{l}
     n\\
     m
   \end{array}\right]_{q^{-1}}
   \exp\left[i(2m-n)sx\right]
   \label{MacqHermite}
\end{eqnarray}
where the $q$-binomial coefficient is defined by
\begin{equation}
   \left[
     \begin{array}{l}
           n\\
           m
     \end{array}
   \right]_{z} =
   \disp\frac{\disp\prod^{n-1}_{k=0}(1-z^{2(k+1)})}
   {\disp\prod^{n-m-1}_{k=0}(1-z^{2(k+1)})\prod^{m-1}_{k=0}(1-z^{2(k+1)})}
\end{equation}
provided that $\prod^{-1}_{k=0}(1-z^{2(k+1)})=1$.
This function $H_n(x;q)$ is nothing but the $q$-Hermite function of
Macfarlane type\cite{5}.

\section{Q-deformed harmonic oscillators of Dubna type ($q=e^{-s^2}$)}
In the $q$-deformed harmonic oscillator of Dubna type, no upper bound
exists in the eigenvalue spectrum in Eq.~(\ref{eigenvaluespec}),
since $q = e^{-s^2} \leq 1$. In sharp contrast to the ordinary harmonic
oscillator where the energy eigenvalue increases with equal spacing,
the spacing of adjacent eigenvalues increases as the power of
$q^{-2} = \exp(2s^2)$, {\it i.e.}, $E_{n+1}-E_n = \exp[2(n+1)s^2]$. 

For the $q$-deformed commutation relation in Eq.~(\ref{q-mutator}) to hold,
the functions $f(x)$, $g(x)$ and $h(x)$ must satisfy the relations
\begin{eqnarray}
    \disp\left[\frac{f(x+2i s)}{f(x+i s)}\right]^2
    = q^{-2}\left[\frac{f(x+i s)}{f(x)}\right]^2
      \exp\{i[h(x)-h(x+2i s)]\},
   \label{Dubnacond1}
\end{eqnarray}
\begin{eqnarray}
   \disp\left[\frac{f(x)}{f(x-i s)}\right]^2
   =q^{-2}\left[\frac{f(x-i s)}{f(x-2i s)}\right]^2
   \exp\{i[h(x)-h(x-2i s)]\}
   \label{Dubnacond2}
\end{eqnarray}
and
\begin{eqnarray}
    \disp q\left[
    \frac{f(x)^2 f(x+i s)^{-2}}
    {g(x+i s)^2}\right.&&\!\!\!\!\!\!\!\!\!\!\!\!\left.
   +\frac{f(x)^2f(x-i s)^{-2}}
   {g(x-i s)^2}\right]
   \nonumber\\
   \noalign{\vspace{-2mm}}\nonumber\\
      &&\hspace{-2cm}\disp -q^{-1}
   \left[\frac{f(x)^{-2}f(x+i s)^2}
   {g(x+i s)^2}
   +\disp\frac{f(x)^{-2}f(x-i s)^2}
   {g(x-i s)^2}\right] = -4s^2 g(x)^2.
   \label{DubnaF3}
\end{eqnarray}
The former two relations in Eqs.~(\ref{Dubnacond1}) and
(\ref{Dubnacond2}) require that the functions $h(x)$ and $f(x)$
being analytic in the complex $x$ plane have to satisfy the
following difference equations
\begin{eqnarray}
    h(x)-h(x+2i s) = l\pi
\end{eqnarray}
with $l\in{\bf Z}$ and
\begin{eqnarray}
   F(x+i s) = q^{-2}\exp{(i l\pi)}F(x).
\end{eqnarray}
General solutions of these equations are given by
\begin{equation}
   h(x) = i \frac{\pi l}{2s}x
   + \sum_{n=-\infty}^\infty a_n \exp{\left(\frac{\pi n}{s}x\right)}
\end{equation}
and
\begin{eqnarray}
   F(x) \equiv \disp{\left[\frac{f(x+i s)}{f(x)}\right]^2
        =\sum_{n=-\infty}^{\infty}b_n \exp\left(
         \frac{2n\pi}{s}x+\frac{l\pi}{s}x-2i sx\right)}.
\end{eqnarray}
The conditions that $h(x)$ is real for $x\in {\bf R}$ and has a definite limit
at $s\rightarrow 0$ specify the function $h(x)$ to be
\begin{equation}
  h(x)=a_0. 
\end{equation}
Just as in the previous section, the function $F(x)$ in the present case
($q=\exp(-s^2)$) is uniquely determined by the condition
$\lim_{s\rightarrow 0}F(x)=1$ as
\begin{eqnarray}
   F(x) \equiv \disp\left[\frac{f(x+i s)}{f(x)}\right]^2
   = q^{-1}\exp(-2i sx) = \exp(s^2 -2i sx).
\end{eqnarray}
This is identical to the function $F(x)$ in Eq.~(\ref{F(x)solution})
which was obtained in the case of $q=\exp(s^2)$. Therefore, we find
that, in both cases of the $q$-deformed oscillators of Macfarlane and
Dubna types, the part-function $f(x)$ takes the same form which is
given in Eq.~(\ref{f(x)solution}).

The remaining equation (\ref{DubnaF3}) is expressed by
\begin{eqnarray}
  \disp F(x)\left[q^{3}g(x+i s)^2-q^{-1}{g(x-i s)^2}\right]
      &\!\!\!\!\! -  \!\!\!\!\!
	    & F(x)^{-1}\left[q^{-3}g(x+i s)^2-qg(x-i s)^2\right]
   \nonumber\\
   \noalign{\vskip 0.2cm}
    & \!\!\!\!\! = \!\!\!\!\! &\disp -4g(x)^2g(x-i s)^2g(x+i s)^2
   \label{Dubnacond3}
\end{eqnarray}
in terms of the function $F(x)$ in Eq.~(\ref{F(x)}). In appendix, 
we solve this nonlinear difference equation with respect to $g(x)^2$
which has to be real for $x\in\bf R$ and obtain the following solution:
\begin{equation}
   g^{\kappa, \lambda}_{\mu, \nu}(x)^2 
   = G^{\kappa, \lambda}_{\mu, \nu}(x)
     \left(\frac{q^{-1}-q}{s^2}\right)^{\frac{1}{2}}\cos sx
   \label{infinitesolutions}
\end{equation}
with
\begin{eqnarray}
     G^{\kappa, \lambda}_{\mu, \nu}(x) = 
     \tanh^\kappa\left[\frac{(2\mu+1)\pi}{2s}x\right]
     \coth^\lambda\left[\frac{(2\nu+1)\pi}{2s}x\right]
\end{eqnarray}
where $\kappa$ and $\lambda$ are arbitray numbers, and $\mu$ and $\nu$
are arbitrary integers. It is essential to recognize that the factors
$G^{\kappa, \lambda}_{\mu, \nu}(x)$ satisfy the relations
\begin{equation}
 G^{\kappa, \lambda}_{\mu, \nu}(x)\exp\left(i s\frac{d}{d x}\right)
 G^{\kappa, \lambda}_{\mu, \nu}(x)=\exp\left(i s\frac{d}{d x}\right).
 \label{Ginv}
\end{equation}
Owing to these relations, the factors $G^{\kappa, \lambda}_{\mu, \nu}(x)$
have no influence on the structure of the ladder operators $A(x)$ and
$A^\da(x)$. Therefore, it is possible to interpret that these factors
are redundant and that the infinite number of solutions in
Eq.~(\ref{infinitesolutions}) can be reduced to the simplest choice
\begin{equation}
   g(x) = \left(\frac{q^{-1}-q}{s^2}\right)^{\frac{1}{4}}
          \sqrt{\cos sx}.
   \label{gDubna}
\end{equation}
Note here again that a phase of the part-function $g(x)$ is
irrelevant, since only the product of this part-function appears
in the ladder operators. With the part-functions $f(x)$, $g(x)$
and $h(x)$ thus obtained, the $x$-dependence of the ladder
operators is locally and naturally fixed.

In the previous section, the equation (\ref{Ada2}) relating the
operators $A^{\da 2}$, $A^\da A$ and $A^\da$ enabled us to derive
the recursion formula of the eigenfunction of the Hamiltonian.
In the case of the $q$-deformed oscillators of Dubna type,
there exists no such relation. Following Mir-Kasimov \cite{10,11},
we introduce here a new operator as
\begin{eqnarray}
  T=\disp{\frac{1}{g(x)}\cosh\left(i s\frac{d}{dx}\right)
  \frac{1}{g(x)}}.
\end{eqnarray}
which also is not influenced by the factor
$G^{\kappa, \lambda}_{\mu, \nu}(x)$ due to the relation
(\ref{Ginv}). Among the operator $T$ and the ladder
operators, there exists the following bilinear relation
\begin{eqnarray}
   T^2 = s^2 q^{-1}\left(A^{\dag}A + \frac{q}{1-q^2}\right).
\end{eqnarray}
Therefore the operator $T^2$ and the Hamiltonian $H_q$ share the common
eigenfunction $\psi_n(x)$ as
\begin{eqnarray}
   T^{2}\psi_n(x) =  s^2\frac{\ q^{-2n}}{1-q^2}\,\psi_n(x).
\end{eqnarray}
As the square-root of $T^2$, the operator $T$ satisfies
\begin{eqnarray}
    \disp{T\psi_n(x) = \pm s
    \left(\frac{\ q^{-2n}}{1-q^2}\,\right)^{\frac{1}{2}}\,\psi_n(x)},
\end{eqnarray}
since $\psi_n(x)$ is naturally assumed to be non-degenerate.
Without loss of generality, the positive eigenvalue of $T$ can be
taken in the following argument. It is straightforward to prove
that there exists the linear relation
\begin{eqnarray}
  T = \frac{s}{2\sin sx}\left[\frac{1}{\sqrt{q}}\exp[i h(x)]A
     +\sqrt{q}\exp\left[-i h(x)\right] A^{\dag}\right]
   \label{TAAdalinear}
\end{eqnarray}
among $T$ and the ladder operators. Applying this relation to
the eigenfunction $\psi_n(x)$, we find the recursion formula
\begin{eqnarray}
  \disp \psi_{n+1}(x)&\!\!\!=\!\!\!&2\left[\frac{1}{1-q^{2(n+1)}}\right]^{\frac{1}{2}}
   \sin sx \,\exp\left[i h(x)\right]\,\psi_n(x)
   \nonumber\\
   \noalign{\vskip 0.2cm}
  &&\disp{-\left\{\frac{(1-q^{2n})^2}
  {\left[1-q^{2n}\right]\left[1-q^{2(n+1)}\right]}\right\}^{\frac{1}{2}}
   \exp[2i h(x)]\,\psi_{n-1}(x)}.
   \label{Dubnapsirecursion}
\end{eqnarray}

In parallel to Eq.~(\ref{MacHermitedef}) for the $q$-deformed oscillator
of Macfarlane type, let us define the $q$-deformed Hermite function
$H_n(x;q)$ of Dubna type by
\begin{eqnarray}
   \disp{\psi_n(x) = K_0f(x)g(x) s^n
   \exp[i n h(x)]
   \prod_{m=0}^{n-1} \left[1-q^{2(m+1)}\right]^{-\frac{1}{2}}\,H_n(x;q)}
   \label{DubnaHermitedef}
\end{eqnarray}
provided that $\disp\prod_{m=0}^{-1}1/\sqrt{1-q^{2(m+1)}}=1$.
Subsequently, the relation (\ref{Dubnapsirecursion}) leads to the recursion
formula
\begin{eqnarray}
   H_{n+1}(x;q)=\disp\frac{2}{s}\sin sx \,H_n(x;q)
   -\frac{1}{s^2}\left(1-q^{2n}\right)H_{n-1}(x;q).
\end{eqnarray}
for the $q$-Hermite function. Likewise to Eq.~(\ref{Macsecondrecursion}),
we find the second recursion formula
\begin{eqnarray}
     2i s\cos sx H_{n+1}(x ; q)&&
    \nonumber \\
    \noalign{\vskip 0.2cm}
     &&\hspace{-3cm}= q^{-n}\left[\exp(2i sx)H_n(x-i sx ; q)
         -\exp(-2i sx)H_n(x+i sx ; q)\right].
    \label{Dubnasecondrecursion}
\end{eqnarray}
From these formulas, we obtain the power series representation \cite{10,14}
\begin{eqnarray}
      H_n(x;q)=\left(\frac{i}{s}\right)^{n}
               \sum_{m=0}^{n}(-1)^{m}
               \left[
                 \begin{array}{l}
                   n\\
                   m
                 \end{array}
                \right]_{q}
                \exp\left[i(2m-n)sx\right].
   \label{DubnaqHermite}
\end{eqnarray}
Note that the $q$-Hermite functions in Eqs.~(\ref{MacqHermite})
and (\ref{DubnaqHermite}) are periodic functions with the same
period $2\pi/s$. 

The simplest choice of the $h(x)$ function which brings forth the
$q$-Hermite function to the ordinary Hermite function at the limit
$q \rightarrow 1$ is to put $h(x)=0$.
In this case the eigenfunctions $\psi_n(x)$ correspond to
those of the Kasimov solutions except for a different choice of
a measure function for the L${}_2$ norm.

\section{Global structure of the operators and state vectors of
$\bq$-deformed oscillator systems}
In the preceding two sections, the component functions $f(x)$, $g(x)$ and
$h(x)$ of the ladder operators are derived locally from the condition that
the $q$-mutator holds locally for each value of the coordinate $x$.
With these constituents, we must determine here the global structure of
the ladder operators and the eigenfunctions of the Hamiltonian operator.

In both oscillator systems of Macfarlane and Dubna types, the function $f(x)$
is represented as an arbitrary superposition of the Gaussian functions with
center at $x=\frac{2\pi}{s}\times$(integers). It is the freedom existing in
the way of superposition that rules the global structure of the ladder
operators and the state vectors. Namely, different global structure of
the system is realized by different choice of the coeffients $c_m$ in
Eq.~(\ref{f(x)solution}). We distinguish two kinds of global structure
as follows:
\smallskip
\begin{itemize}
\item\ {\bf Aperiodic structure} \\ 
The domain of the oscillator coordinate $x$ is identified directly
with the infinite interval $I_\infty = (-\infty, \infty)$.
The part-function $f(x)$ is defined over the whole interval
$I_\infty$ by the superposition
\begin{equation}
   f(x) = \sum^{\infty}_{m=-\infty}c_m(s) f_m(x)
\end{equation}
of the Gaussian functions
\begin{equation}
  f_m(x)=\exp\left[-\frac{1}{2}\left(x-\frac{2m\pi}{s}\right)^2\right]
  \label{fm}
\end{equation}
with the coefficients $c_m(s)$ which make the ground-state wave
function $\psi_0(x)$ in Eq.~(\ref{groundstate}) to be square-integrable.
For the definite choice of the function $f(x)$, the ladder operators
are defined in Eqs.~(\ref{A}) and (\ref{Adag}) and the eigenvalue
problem of the Hamiltonian consisting of those ladder operators is solved.
There are systems with finite and infinite numbers of Gaussian factors.
Note that the uniform coefficients, {\it i.e.}, $c_m$ = constant
for all $m$, are forbidden. Therefore, any choice of $f(x)$ of this
kind over the whole interval $I_\infty$ can not preserve the periodic
character carried intrinsically by the $g(x)$ functions
in Eqs.~(\ref{gMac}) and (\ref{gDubna}) and the $q$-Hermite
functions in Eqs.~(\ref{MacqHermite}) and (\ref{DubnaqHermite}).
\smallskip
\item\ {\bf Periodic structure} \\
The domain of the oscillator coordinate is considered to be
covered by the infinite sum of the finite interval as
\begin{equation}
   I_\infty = \bigcup^\infty_{m=-\infty}\,I_m
\end{equation}
where
\begin{equation}
  I_m = \left[\frac{(2m-1)\pi}{s}, \frac{(2m+1)\pi}{s}\right].
\end{equation}
In the defining equations (\ref{A}) and (\ref{Adag}),
the ladder operators $A(x)$ and $A^\da(x)$ are constructed
sectionally on each interval $I_m$ with the part-function
$f_m(x)$ in Eq.~(\ref{fm}). All operators and functions
including the part-function $f(x)$ have to be defined in this way.
Then all of them are smoothly connected over the whole interval
$I_\infty$. As a result, the $q$-deformed oscillator system
thus specified becomes periodic in conformity with the property
of the part-functions $g(x)$ and the $q$-Hermite functions.
\end{itemize}
\smallskip
Both of the $q$-deformed harmonic oscillators of Macfarlane and
Dubna types can have these global structure. Namely,
in addition to the periodic solution which was already studied
intensively\cite{5,6}, there exist an infinite variety of
aperiodic solutions in the $q$-deformed oscillators of both types.

All eigenfunctions $\psi_n(x)$ in Eqs.~(\ref{MacHermitedef})
and (\ref{DubnaHermitedef}) are proportional to the part-function
$g(x)$. This is essential in the oscillator system of Dubna type.
This important characteristic arises from the fact that all
eigenfunctions include the product of the part-function $f(x)$ and
the q-Hermite function. The mechanism which creats the part-function
$g(x)$ being proportional to $\cos sx$ is explicitly embodied in
the second recursion formula (\ref{Dubnasecondrecursion}).

Therefore, every elements in the Hilbert space ${\cal H}$
generated by the eigenfunctions $\psi_n(x)$ are considered to include
the part-function $g(x)$. In the oscillator system of Dubna type,
the ladder operators being proportional inversely to $g(x)$ have
singularities at $x=\frac{\pi}{2s}\times({\rm odd\ number})$.
These singularities are cancelled by the zero-points of the elements
of the Hilbert space ${\cal H}$ in the inner product. Owing to
this cancellation, the raising operator $A^\da(x)$ is proved to be
adjoint to the lowering operator $A(x)$ in the Hilbert space ${\cal H}$,

\section{Discussion}
In this constructive approach, the $q$-deformed harmonic oscillators
of Macfarlane and Dubna types were proved to appear as coordinate
representations of the same algebra of the $q$-deformed ladder
operators, respectively, for disconnected sectors $q > 1$ and
$q < 1$ of the deformation parameter $q$. The eigenvalues of
the Hamiltonian given by the $q$-deformed anti-commutator constitute
the spectrum expressed by the common single functions $E_n(q)$ of
the parameter $q$ irrespective of the choice of representations.
Namely, the single function $E_n(q)$ realizes the energy spectra of
the oscillatoers of Macfarlane type in the $q>1$ sector and that
of Dubna type in the $q<1$ sector.

In sections 4 and 5, the part-functions of the ladder operators
were obtained for respective types of oscillators.
The part-function $f(x)$ turned out to be exactly common in both
types. Furthermore, the other part-functions $g(x)$ and $h(x)$
which are seemingly different can be unified by using parametric
representations as follows:
\begin{equation}
    g(x) = \left(\frac{e^{s^2}-e^{-s^2}}{s^2}\right)^\frac{1}{4}
           \sqrt{\cos tx}
\end{equation}
and
\begin{equation}
    h(x) = -2(s+t)x + a_0.
\end{equation}
This sort of unification is possible, since all the differences
between the two types of oscillators stem simply from
the different choices of the $s$ and $t$ parametrization in
the part-function $D$ of the momentum operator.
Therefore, it is not unnatural and irrelevant to expect that
the eigenfunctions derived in those sections have generic unified
parametrizations. In fact, the eigenfunctions $\psi_n(x)$ in
Eqs.~(\ref{MacHermitedef}) and (\ref{DubnaHermitedef})
have the integrated form
\begin{eqnarray}
  \psi_n(x)=K_0f(x)g(x)s^n&\!\!\!\!\!\!\!\!\!\!\!&\exp\left\{i n[h(x)+(s+t)x]\right\}
 \nonumber\\
 \noalign{\vskip 0.2cm}
  &\!\!\!\!\!\!\!\!\!\!\!&\times\prod_{m=0}^{n-1}
            \left\{e^{(s+t)^2}[1-e^{-2s^2(m+1)}]\right\}^{-\frac{1}{2}}
            H_n(x;e^{-s^2})
\end{eqnarray}
with the unified $q$-Hermite function $H_n(x;e^{-s^2})$ which satisfies the
recursion formulas
\begin{eqnarray}
  H_{n+1}(x;e^{-s^2})=\disp\frac{i}{s}
 \left(e^{\frac{1}{2}(s+t)^2-i sx}-e^{-\frac{1}{2}(s+t)^2+i sx}\right)
 H_{n}(x;e^{-s^2})
 \nonumber\\
 \noalign{\vskip 0.2cm}
  -\frac{1}{s^2}(1-e^{-2ns^2})H_{n-1}(x;e^{-s^2})
\end{eqnarray}
and
\begin{eqnarray}
  i s\left\{\exp\left[i sx -\frac{1}{2} (s+t)^2\right]
               +\exp\left[i sx -\frac{1}{2} (s+t)^2\right]\right\}
               \,H_{n+1}(x;e^{-s^2})
 \nonumber \\
 \noalign{\vskip 0.2cm}
 = e^{-ns^2}\left\{\exp\left[2i sx -(s+t)^2\right]H_n(x-i s;e^{-s^2})\right.
 \nonumber \\
 \noalign{\vskip 0.2cm}
         \left.- \exp\left[-2i sx +(s+t)^2\right]H_n(x+i s;e^{-s^2})\right\}.
\end{eqnarray}
The power series representation of the unified $q$-Hermite function is given by
\begin{eqnarray}
   \hspace{-1cm}H_{n}(x;e^{-s^2})=\disp\left(\frac{i}{s}\right)^n
  \sum^{n}_{m=0}(-1)^m
  \left[\begin{array}{l}n\\m\end{array}\right]_{e^{-s^2}}
  \exp\{(2m-n)[i sx-\frac{1}{2} (s+t)^2]\}.
\end{eqnarray}
The parametric unification realized in this way is a direct proof of
close kinship of two types of the $q$-deformed oscillator systems.

Main difference between two types of the $q$-deformed oscillators appears in
the part-function $g(x)$. However, since the ladder operators are more basic
than the individual part-functions, we realize that those differences are
rather superficial and not so essential. In fact, the singularities arising
from the factor $1/g(x)$ in the ladder operators do not cause any harm in
their action and the redundant factors investigated in the appendix have no influence to the ladder
operators themselves.

The global structure considered in the previous section is also a generic
character for the $q$-deformed harmonic oscillators investigated in the present
formalism. Existence of the periodic solution and the infinite numbers of
aperiodic solutions shows a rich structure in our $q$-deformed oscillator
system. 

\section*{Appendix}
\appendix
Let us solve the difference equation (\ref{Dubnacond3}) with
respect to $g(x)^2$. Noting that the function $F(x)$ is
proportional to the factor $\exp\left(-2i sx\right)$, we
set
\begin{eqnarray}
  g(x)^2=\xi(x)\exp\left(-i s x\right)
        +\eta(x)\exp\left(i s x\right)
\end{eqnarray}
in which $\xi(x)$ and $\eta(x)$ are assumed not to include the
factor $\disp\exp\left(i sx\right)$. Substitution of this
expression into the equation (\ref{Dubnacond3}) results in the
following equations for unknown functions $\xi(x)$ and $\eta(x)$ as
\begin{eqnarray}
  	\disp\frac{1}{q}\left[q^2\xi(x+i s)-\xi(x-i s)\right]
     =-4s^2 \xi(x)\xi(x-i s)\xi(x+i s),
\end{eqnarray}
\begin{eqnarray}
     \disp q \!\!\!\!\!\!\!\!\!\!&&\left[q^2\xi(x+i s)-q^{-4}\xi(x-i s)\right]
   \nonumber\\
   &&\quad\quad=-4s^2 \left[\xi(x)\eta(x-i s)\eta(x+i s)
 +q^{2}\eta(x)\xi(x-i s)\eta(x+i s)\right.
   \nonumber\\
   \!\!\!\!\!\!\!\!\!\!&&\hspace{5cm}+\left.q^{-2}\eta(x)\eta(x-i s)\xi(x+i s)\right],
\end{eqnarray}
\begin{eqnarray}
     \disp\frac{1}{q}\disp\!\!\!\!\!\!\!\!\!\!&&\left[q^4\eta(x+i s)-q^{-2}\eta(x-i s)\right]
     \nonumber\\
  \!\!\!\!\!\!\!\!\!\!&&\quad=-4s^2 \left[\eta(x)\xi(x-i s)\xi(x+i s)
     +q^{2}\xi(x)\xi(x-i s)\eta(x+i s)\right.
  \nonumber\\
   &&\hspace{5cm}+\left.q^{-2}\xi(x)\eta(x-i s)\xi(x+i s)\right]
\end{eqnarray}
and
\begin{eqnarray}
     \disp q\left[\eta(x-i s)-q^{-2}\eta(x+i s)\right]
     =-4s^2 \eta(x)\eta(x-i s)\eta(x+i s).
\end{eqnarray}
These simultaneous difference equations have the solutions
\begin{eqnarray}
   \xi(x)=\eta(x)=\frac{1}{2}\disp\left(\frac{q^{-1}-q}{s^2}
   \right)^\frac{1}{2}
   \tanh^\kappa\left[\frac{(2\mu+1)\pi}{2s}x\right]
   \coth^\lambda\left[\frac{(2\nu+1)\pi}{2s}x\right]
\end{eqnarray}
where $\kappa$ and $\lambda$ are arbitray numbers, and $\mu$
and $\nu$ are arbitrary integers. Consequently we find the solutions of
the equation (\ref{Dubnacond3}) as
\begin{eqnarray}
  g_{\mu,\nu}^{\kappa,\lambda}(x)^2=G^{\kappa, \lambda}_{\mu, \nu}(x)
  \disp\left(\frac{q^{-1}-q}{s^2}\right)^\frac{1}{2}
  \frac{1}{2}\left[\exp\left(-i sx\right)\pm \exp\left(i sx\right)\right]
\end{eqnarray}
where
\begin{eqnarray}
     G^{\kappa, \lambda}_{\mu, \nu}(x) = 
     \tanh^\kappa\left[\frac{(2\mu+1)\pi}{2s}x\right]
     \coth^\lambda\left[\frac{(2\nu+1)\pi}{2s}x\right].
\end{eqnarray}
The condition that $g(x)^2$ has to be real for $x\in\bf R$
selects out the solutions
\begin{eqnarray}
  g_{\mu,\nu}^{\kappa,\lambda}(x)^2
  =\disp\left(\frac{q^{-1}-q}{s^2}\right)^\frac{1}{2}
   G^{\kappa, \lambda}_{\mu, \nu}(x)\cos sx.
\end{eqnarray}

\end{document}